# Magnetic resonance in nanoparticles: between ferro- and paramagnetism

N. Noginova[1], F. Chen[1], T. Weaver[1], E. P. Giannelis[2], A. B. Bourlinos[2], and V.A. Atsarkin[3]

[1]*Norfolk State University, Norfolk VA USA*
[2] *Cornell University, Ithaca, NY, USA*
[3]*Institute of Radio Engineering & Electronics, Moscow 125009, Russia*

Magnetic nanoparticles of γ-$Fe_2O_3$ coated by organic molecules and suspended in liquid and solid matrices, as well as a non-diluted magnetic fluid have been studied by electron magnetic resonance (EMR) at 77-380 K. Slightly asymmetric spectra observed at room temperature become much broader, symmetric, and shift to lower fields upon cooling. An additional narrow spectral component (with the line-width of 30 G) is found in the diluted samples, its magnitude obeying the Arrhenius law with the activation temperature of about 850 K. The longitudinal spin-relaxation time, $T_1 \approx 10$ ns, was determined by the specially developed modulation method. Angular dependence of the EMR signal position in field-freezing samples unambiguously points to the domination of the uniaxial magnetic anisotropy. Substantial alignment is achieved in moderate freezing fields of 4-5 kG, suggesting formation of dipolar-coupled chains consisting from several particles separated by organic nanolayers.

The shift and broadening of the spectrum upon cooling are ascribed to the role of the surface layer, which is considered with taking into account the strong surface–related anisotropy. To describe the overall spectrum shape, a "quantization" model is used which includes summation of the resonances corresponding to various orientations of the particle's magnetic moment at a given temperature. This approach, supplemented with some phenomenological assumptions, provides satisfactory agreement with the experimental data.

## I. Introduction

Magnetic nanoparticles attract considerable interest due to their unusual magnetic properties and many technological applications, such as in nanoscale engineering, catalysis, mineralogy, biology, and medicine (for a review, see Refs. [1,2]). There is a very important and interesting fundamental issue as well: nanometer-scale magnetic objects are at the interface between quantum dynamics of few interacting spins and many-particle behavior commonly described in terms of classical thermodynamics. The gap between dynamical (reversible) and thermodynamical (irreversible) approaches represents one of the most general problems in physics. Thus, studying magnetic nanoparticles, which could be considered as intermediate case between these two areas, one can get essential information on this intriguing problem.

Among many publications on magnetic nanoparticles, there are a considerable number of studies performed by means of electron magnetic resonance (EMR). The theory of magnetic resonance in such superparamagnetic systems has been developed in Refs. [3-5], based on the phenomenological equation of motion for a classical magnetic moment **μ** under conditions of ferromagnetic resonance (FMR). The main result obtained in this theory is a sort of averaging caused by thermal re-orientations of the magnetic moment and leading to the effective reduction of the anisotropy field with increase in temperature.

A number of magnetic resonance experiments were performed by various authors on assemblies of randomly oriented nanoparticles (as a rule, nanoparticles were embedded in a diamagnetic matrix to weaken or even exclude the inter-particle interaction) [6-19]. For the most part the agreement between experimental data and theoretical predictions is rather poor and does not provide an opportunity of accurate quantitative analysis of the experimental results. The only exception is the high-temperature limit where, according to the theory, the spectrum progressively collapses into a single, nearly Lorentzian line. At lower temperatures, the specific pattern, predicted by the theory, was not observed; instead, a significant broadening of the single line was found together with its progressive shifting to lower fields with decrease in temperature.

Thus, further studies and interpretation of the magnetic resonance experimental data in assemblies of magnetic nanoparticles remain very important. Such investigation, supported by a theoretical approach combining ferromagnetic (classical) and paramagnetic (quantum) considerations, is the main purpose of this work.

**II. Experimental Techniques and Results.**

The experimental samples were prepared using the solvent-free ferrofluid containing surface functionalized maghemite ($\gamma$-$Fe_2O_3$) nanoparticles of d~5 nm diameter. Such ferrofluid has been produced by attaching a corona of flexible chains onto maghemite nanoparticles. Specifically, reaction of a positively charged organosilane (($CH_3O)_3Si(CH_2)_3N^+(CH_3)(C_{10}H_{21})_2Cl^-$) with surface hydroxyl groups on the nanoparticles leads to a permanent covalent attachment to the surface and renders the nanoparticles positively charged. A counter anion is present to balance the charge leading to a hybrid nanoparticle salt. The sulfonate anions ($R(OCH_2CH_2)_7O(CH_2)_3SO_3^-$, R: $C_{13}$-$C_{15}$ alkyl chain) were used, yielding a liquid at room temperature with a nanoparticle content around 40 wt %. [20]. TEM picture and the schematic of the nanoparticles are shown in Fig. 1.

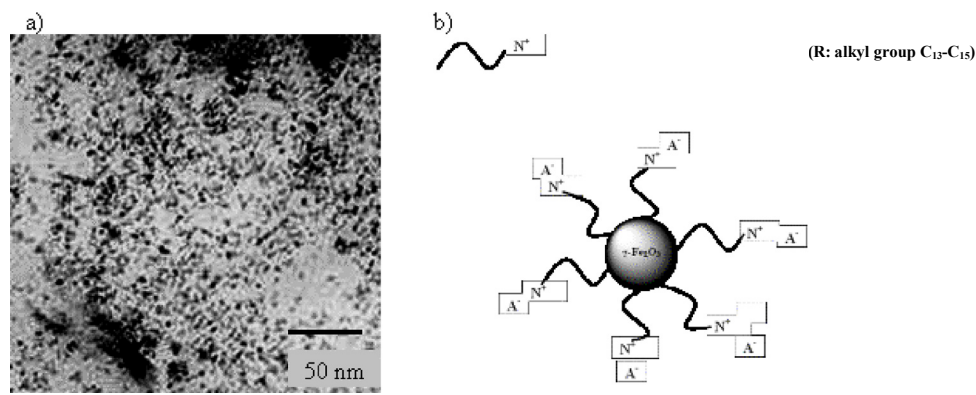

Fig. 1. a: TEM picture of the ferrofluid; b: Schematic of surface functionalized $\gamma$-$Fe_2O_3$ nanoparticles.

The size distribution of the nanoparticles was found to be nearly log-normal, with the mean diameter of 4.8 nm and dispersion $\sigma$ = 0.15. Apart from well separated nanoparticles, large clusters (aggregates) are also seen at the TEM picture. To prepare experimental samples with different concentrations, the nanoparticles were dispersed in liquid (toluene) and solid (polystyrene) matrices.

The electron magnetic resonance studies were performed using EPR Spectrometer Bruker EMX operating at 9.8 GHz (X band); modulation frequency was 100 kHz. The commercial gas-flow cryostat was used which allowed to achieve temperatures in the range of 77÷360K inside the quartz dewar tube placed into the microwave cavity. The cavity itself was kept at room temperature, and its quality factor was not changed upon cooling. At each temperature value, the magnitude of the resonance signal under study was calibrated versus the reference sample (MgO:$Mn^{2+}$) situated outside the dewar tube.

The longitudinal relaxation time ($T_1$) was measured by modulation technique using a home-made apparatus with detection of longitudinal magnetization component oscillating at modulation frequency 1.6 MHz [21]. Both the "phase" and "amplitude" versions were employed; in the latter case, the diphenylpicrylhydrazyl (DPPH) was used as a reference sample.

Typical EMR signals at room temperature (T=295K) are shown in Fig. 2. The signal can be described approximately as a sum of two lines, the broad one with the peak-to-peak width of about 500 G, and the narrow one with the width of about 30 G. The narrow line has the g-factor, g ≈2; it can be seen more clearly in the derivative of EMR signal (that is the second derivative of the EMR absorption), see inset in Fig. 2. This narrow signal is observed in well-diluted samples; with increasing nanoparticle concentration it becomes broader, lower in amplitude, and not well resolved.

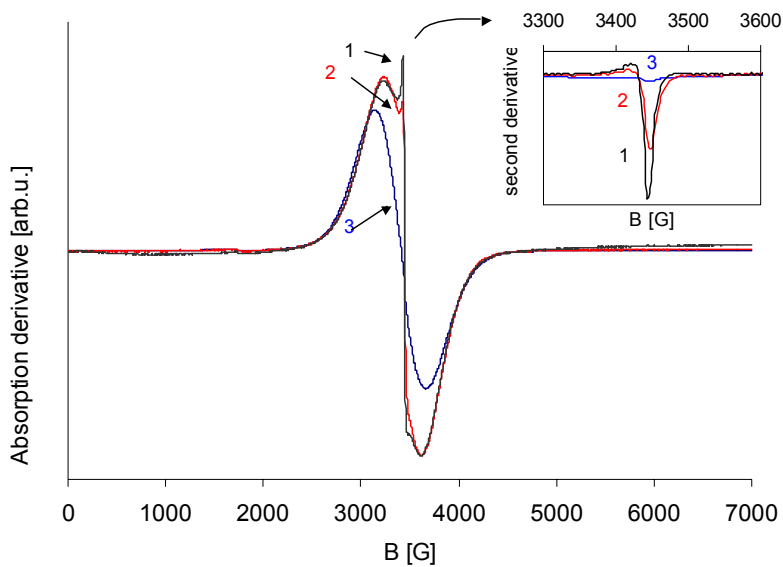

Fig. 2. EMR signal in the ferrofluid diluted in toluene at T=295 K. The degree of dilution (by weight) is 1:6000 (1), 1:200 (2), and 1:6 (3). The spectra are normalized to the same intensity (double integrated area). Inset: The derivative of the EMR signal demonstrating the narrow peak.

The EMR spectra taken at different temperatures are shown in Fig. 3. One can see that, with the decrease in temperature, the broad signal shifts toward lower fields, its width increases.

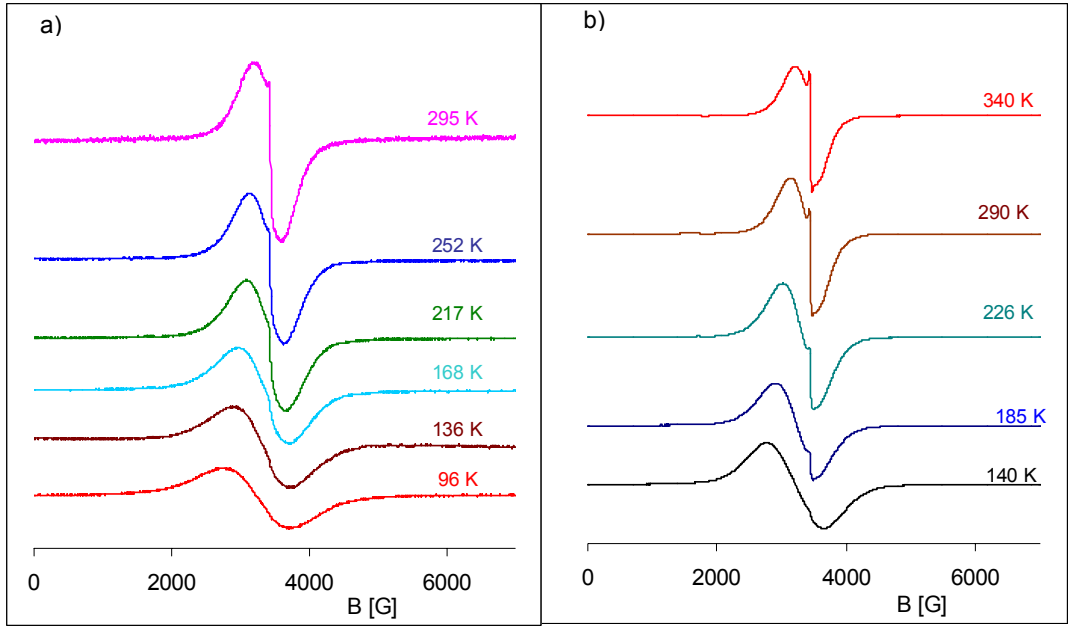

Fig. 3. EMR in γ-Fe$_2$O$_3$ suspensions in the polymer matrix (a) and toluene (b) at different temperatures.

The temperature dependencies of the line shift, $B_0-B_0(T)$, and peak-to-peak width $\Delta B_{pp}$ for the γ-Fe$_2$O$_3$ suspension in the polymer matrix are presented in Fig. 4(a). Here $B_0(T)$ is the resonance field determined as the point where the absorption derivative equals zero at the temperature T, and $B_0$=3442 G represents its asymptotic value at high temperatures. The integrated EMR intensity, $I_{EMR}$, calculated through double integration of the absorption-derivative spectrum is shown in Fig. 4(b). In both diluted and dense ferrofluid, the EMR intensity follows the same temperature dependence: as the temperature decreases, $I_{EMR}$ increases and then saturates to some constant value.

We made an attempt to find an evidence for ferromagnetic blocking frequently observed in magnetic nanoparticles at low enough temperatures (see, for example, Refs. [1, 12, 17, 22, 23]). The EMR intensity has been measured under conditions of zero field cooling (ZFC). Each point in the temperature dependence was obtained after prior heating the sample up to 300 K and subsequent cooling to the desired T value at B=0 and taking the EMR spectrum during the first up-field sweep. No difference in the EMR intensity was found in comparison to the field cooling (FC) procedure, indicating the absence of the blocking effect in our experiments. Evidently, this can be explained by the influence of the external field B ~ 3 kG which strongly exceeds the effective anisotropy field $B_a$ of the particle. Note that the blocking of maghemite nanoparticles in such magnetic fields was observed previously only at T<40 K [12, 23].

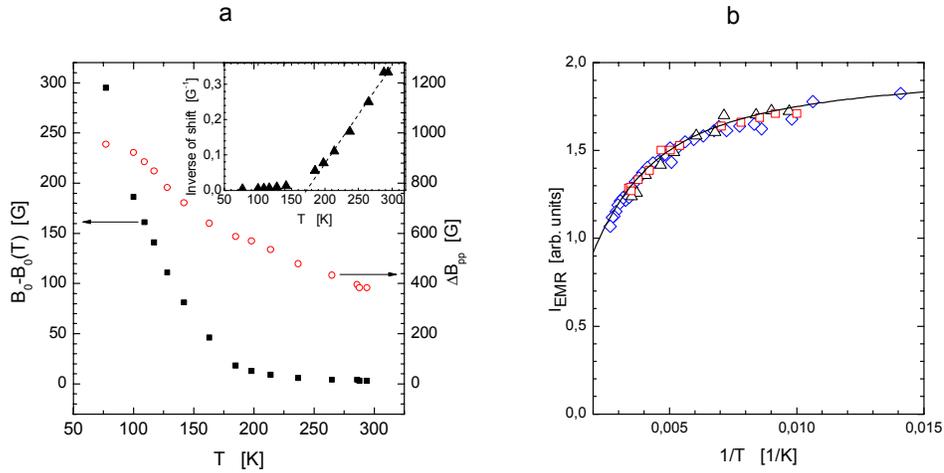

Fig. 4. a: Temperature dependencies of the EMR line shift (filled squares, left scale) and width (open circles, right scale) for the γ-Fe$_2$O$_3$ suspension in the polymer matrix. Inset: the inverse of the shift, $[B_0 - B_0(T)]^{-1}$, *versus* T. The dashed line represents the Curie-Weiss dependence.

b: EMR intensity in nanoparticle systems diluted in polymer (squares), diluted in toluene (triangles) and ferrofluid as it is (diamonds). The data obtained with different samples are normalized to the same value at 295 K. Solid line is the model with $\mu B_0/k_B = 800$ K, see Eq. (1).

Opposite to the broad signal behavior, the narrow signal remains at the same field; its amplitude decreases steeply with decrease in temperature, see Fig. 5(a). The shape of the narrow component (both in the first derivative and second derivative presentations) was thoroughly analysed and found to be temperature independent, at least in its central (peak-to peak) part. This finding enables one to consider the data presented in Fig. 5(a) as the temperature dependence of the intensity.

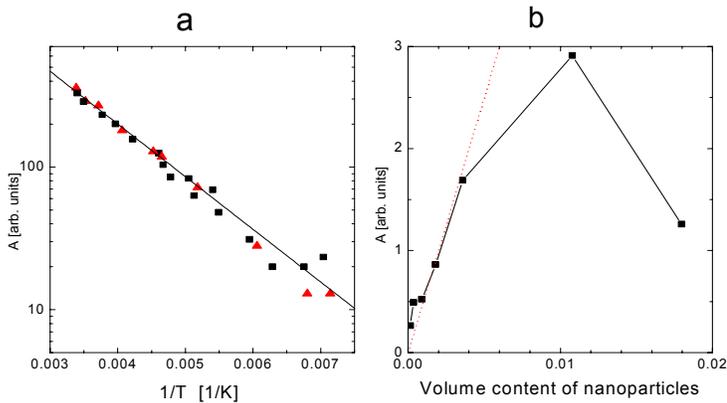

Fig. 5. Properties of the narrow spectral component.
a: the magnitude of the narrow component in polymer (squares) and toluene (triangles) matrices. Solid line is $\exp(-E_a/k_BT)$, with $E_a/k_B = 850$ K.

b: the magnitude of the narrow line *versus* nanoparticle concentration in toluene (T=295K). Solid line connects the experimental points. Dotted line shows linear dependence at low concentration.

To check whether the narrow component was related to the organic groups on the nanoparticle surface, we ran the ESR spectra of the systems containing similar surface modified silica nanoparticles, fabricated by the same method [20]. No signal has been observed. Note that such a double-feature spectra were previously reported for the dispersed maghemite nanoparticles [9, 11], as well as various superparamagnetic and exchange-coupled clusters in solid matrices [13, 14].

The dependence of the narrow line peak magnitude on the nanoparticle volume content ($c$) in liquid toluene solutions is shown in. Fig. 5(b). All data are normalized to the sample volume. At low concentrations ($c \leq 4 \cdot 10^{-3}$), the peak magnitude is proportional to $c$, whereas the pronounced drop occurs at $c > 0.01$, accompanied by the 10-20% increase of the peak-to-peak line-width (not shown in the Figure). In the concentrated ferrofluid, the narrow feature is not observable.

The longitudinal relaxation time $T_1$ corresponding to the observed EMR spectrum was measured by the modulated technique with longitudinal detection [21]. The value of $T_1 = (10\pm3)$ ns was obtained for the concentrated samples in the temperature range of 77- 300 K.

Following the method suggested in Ref. [8], we study the effect of field-freezing. Samples diluted in toluene (freezing temperature $T_{fr}$ = 180 K) have been cooled and frozen in the external magnetic field $\mathbf{B_{fr}}$. We compare the EMR of the field frozen (FF) and zero field frozen (ZFF) samples, and measure the field and angular dependences of the position of the broad signal (determined as a point where the signal crosses zero). As one can see (Fig. 6), the EMR in FF sample is shifted from the position of the ZFF experiment toward lower field if the measuring field, $\mathbf{B_m}$, is parallel to $\mathbf{B_{fr}}$ (the angle, β, between $B_m$ and $B_{fr}$ equals zero) or toward higher field at the perpendicular orientation (β= 90 deg.).

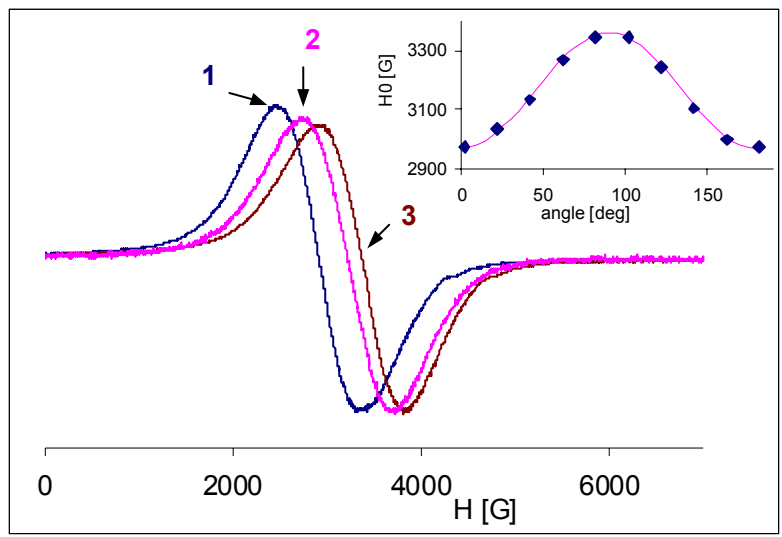

Fig. 6. EMR at 77 K in samples diluted in toluene after ZFF (Trace 2) and FF ($B_{fr}$= 7 kG). Traces 1 and 3 correspond to the orientations of the measuring field β = 0 and 90 degree correspondingly. Inset: the line position in the dependence on the orientation of the measuring field **$B_m$** relative to the direction of the freezing field **$B_{fr}$**. Dots are experiment, solid trace corresponds to $\cos^2\beta$.

This shift depends on the freezing field, demonstrating gradual saturation for fields higher than 4 kG, see Fig. 7. No FF effect was observed in nanoparticles dispersed in solid polymer matrix.

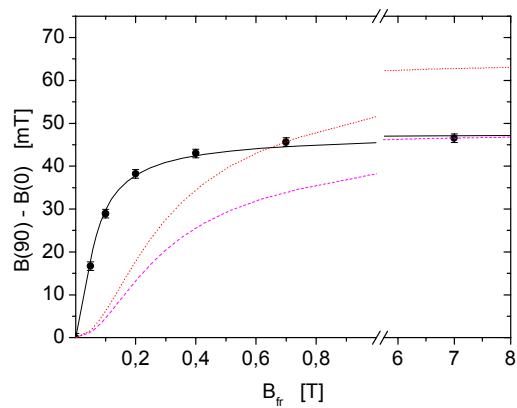

Fig. 7. The difference between resonance fields $B_0(\beta=90$ deg.) and $B_0(\beta=0)$ after the FF procedure *versus* the freezing field $B_{fr}$. Black squares: experiment at 77 K; curves: calculation based on Eqs. (22)-(24) with η=4300 K, $B_a$=800 G (solid); η=800 K, $B_a$=2040 G (dotted); and η=800 K, $B_a$=2380 G (dashed).

The orientation dependence of the line position demonstrates maximum at 90° and can be fitted well with the dependence $\cos^2\beta$, (see the inset in Fig. 6) similar to results obtained in Ref. [8]. According to [8], it indicates predomination of the axial type of magnetic anisotropy (in the case of the cubic magnetocrystalline anisotropy, one would observe the dependence with two maxima). This issue will be discussed below (Section III.F).

**III. Theory and discussion**

**A. General remarks**

The samples under study are assemblies of small (d ≈ 5nm) particles of $\gamma$-$Fe_2O_3$ embedded into non-magnetic matrix (liquid or solid). The Curie temperature of bulk maghemite, $(T_C)_{bulk}$=860 K [8], is higher than the measurement temperatures (77-380 K), and the nanoparticles are formally in the ferromagnetic single-domain state. The anisotropy field $B_a$ can be considered to be much lower than the external field B under the EMR conditions. In this case, the direction of the effective field $\mathbf{B_e}$ practically coincides with direction of $\mathbf{B}$, and, in the ground state, all individual magnetic moments $\boldsymbol{\mu}$ are aligned along the same direction. This entirely polarized state is, however, disturbed by thermal fluctuations. Magnetic energy of a nanoparticle, $U_m$ = - ($\boldsymbol{\mu} \cdot \mathbf{B}$), is comparable with $k_BT$ (where $k_B$ is the Boltzmann constant); as a result the direction of $\boldsymbol{\mu}$ fluctuates, giving rise to specific superparamagnetic properties.

The theory of magnetic dynamics and EMR in superparamagnetic objects was developed in Refs. [3-5]. The version allowing direct comparison with experimental data was worked out by Raikher and Stepanov (RS) [3,4]. Starting from the Landau-Lifshits (LL) equation, they took into account thermal re-orientations of $\boldsymbol{\mu}$ describing it in terms of rotary diffusion, and using the approximation $B_a$<<B, obtained the expression for the magnetic resonance absorption. The main result of the RS theory is the "dressing" of the anisotropy field: in fact, the random walking of $\boldsymbol{\mu}$ over all possible orientations leads to averaging and effective reducing $B_a$.

As a result, the EMR spectrum in superparamagnetic region strongly depends on temperature. In the low-temperature limit ($\xi \equiv \mu B/k_BT$ >>1), random distribution of the anisotropy axes results in huge inhomogeneous broadening (with overall width of $1.5B_a$). The theory [3, 4] predicts for this case a specific, highly asymmetric line shape characteristic of ferromagnetic resonance (FMR) on powdered samples. As temperature increases, the width and asymmetry of the spectrum decreases progressively. Finally, the EMR spectrum is expected to collapse into a single Lorentzian line centered at $B_0=\omega/\gamma$, where $\omega$ is the operating frequency and $\gamma$ the gyromagnetic ratio. At further heating, this line broadens due to the increase of the relaxation rate. Numerically calculated spectra for a set of typical parameters are presented in Refs. [3, 4].

However, much of the reported experimental data hardly agree with these predictions. In particular, the EMR spectra of maghemite nanoparticles of a various size and in various matrices were found to be different from the RS theoretical calculations, especially at low temperatures. The same is true for our data as well: as it is seen from Fig 3, the spectra become broader and more symmetric as temperature decreases, in contradiction with the RS model. Besides, the line is shifted to lower fields upon cooling. Finally, an additional narrow line is observed at B≈$B_0$, with the magnitude strongly dependent on temperature. Similar features were reported previously for many nanoparticle systems, see for example Refs. [6-19]. Various explanation of these "anomalies" have been proposed; their basic idea is to account for specific features of the spins disposed on the particle surface. In the following sub-sections, we will discuss these issues, as well as the FF phenomena and effect of inter-particle interactions in more details.

**B. Estimating the nanoparticle magnetic moment from EMR spectra**

One of the most important quantities needed for interpretation of the experimental data is the magnetic moment μ of an individual nanoparticle. In principle, this value could be estimated as μ = $VM_s$, where V is the particle volume and $M_s$ is its saturated magnetization. However, $M_s$ in nanoparticles can deviate from its bulk value of $4 \cdot 10^5$ A/m (about 80 e.m.u./g) [17]. The EMR spectra provide an opportunity to estimate μ directly from the temperature dependence of the EMR intensity.

As it was mentioned above, nanoparticles dispersed in a non-magnetic matrix can be considered as superparamagnetic objects with a very large spin S~$10^3$-$10^4$; the temperature dependence of their static magnetic susceptibility obeys the formula:

$$\chi(T) = C \cdot L(\xi) \qquad (1)$$

where C is a temperature-independent coefficient, and L(ξ) = coth(ξ)-1/ξ is the Langevin function. Under EMR conditions, the specific form of the χ(T) dependence is determined mainly by the value of μ. Thus, taking into account that the susceptibility is proportional to the EMR intensity (double integrated area under the EMR line), one can determine the μ value from the experimental data, Fig. 4(b).

The best fit of the experimental dependence with Eq.(1) is obtained at η = 800 K, where η ≡ $\mu B_0/k_B$ is introduced, with $B_0$=3.44 kG. For $M_s$ =$4 \cdot 10^5$ A/m, this corresponds to a nanoparticle having diameter of 5.5 nm which is close to the values of *d* estimated from TEM studies.

It is instructive to compare the χ(T) dependence shown in Fig. 4(b) with that obtained on similar nanoparticles by static magnetization method [17, 24]. In the latter case, a small term linear in ξ was observed in addition to Eq.(1), suggesting the contribution of the surface layer which remains paramagnetic down to low temperatures. The absence of such term in our EMR experiments can be explained by strong anisotropy experienced by the surface spins which shifts their resonance frequency beyond the range available with our spectrometer (see the next Section).

**C. Effect of the surface**

The surface effects in magnetic nanoparticles are intensively discussed in the literature. The specific surface-related magnetic anisotropy, which combines strong anisotropy on the particle surface with its deviation from the spherical shape, was introduced by Néel [25]. This mechanism was used recently by Gazeau et al. [8] to explain the large uniaxial anisotropy observed in the field freezing experiments in maghemite nanoparticles. However, a nearly symmetric shape of the EMR spectrum, especially at low temperatures (Fig. 3), suggests a sufficiently small anisotropy field $B_a$ for an individual nanoparticle, not exceeding the observed line-width. We believe that the large values for the axial anisotropy estimated in Ref. [8] are primarily caused by the collective effects. These effects are discussed below in Section III.F. Here we focus on other characteristic features of the observed spectra, such as the shift and broadening.

The shift of the EMR spectrum to lower magnetic fields upon cooling is typical for magnetic nanoparticles (see, for example, Refs. [2, 8, 12, 15, 24]). As a rule, this shift is assigned to surface phenomena, particularly, to the "exchange anisotropy" arising at the interface between ferro- and antiferromagnetic (or spin-glass) layers [26, 27]. In this context,

Kodama and Berkowitz [28] performed sophisticated model calculations for the $\gamma$-$Fe_2O_3$ nanoparticles and proved the spin-glass-like arrangement of the surface spins. Some experimental evidences for the surface spin glass were reported in Refs. [12, 23]. This model is restricted, however, to the low-temperature behavior and related to the problem of the ground state; besides, it deals with static magnetization rather than magnetic resonance. In what follows, we make an effort to suggest a simpler approach applicable to relatively high temperatures and spherical shape without any distortion. In this approach, the bulk and the surface are considered separately with the effect of their interaction introduced semi-phenomenologically.

Consider a spherical magnetic nanoparticle of the radius R subjected to an external magnetic field **B**||z (Fig. 8). The spin-Hamiltonian of the whole spin system reads:

$$H = H_v + H_s + H_{sv} \qquad (2)$$

where $H_v$ and $H_s$ are the energy operators of the surface atomic layer and the rest volume, respectively, and $H_{sv}$ describes the interaction between them. Suppose first that the condition of

$$\gamma B_{as} > J_{sv} \qquad (3)$$

is fulfilled for the surface spins, where $B_{as}$ is the surface anisotropy field caused by the electric field gradient (EFG) and directed at the right angle to the surface, and $J_{sv}$ is the exchange interaction (in frequency units) between the surface and bulk spins. The similar inequality was adopted by Dimitrov and Wysin in their computer modeling of the ferromagnetic nanosphere [29]. The EFG is resulted from the lack of the outer oxygen in the first co-ordination octahedron surrounding the magnetic ion ($Fe^{3+}$, for example). Our estimations show that the anisotropy parameter on the surface can exceed that in the bulk by about two orders of magnitude. In such a case, the magnitude of "zero-field splitting" caused by the anisotropy field may be of the order of $10^2$-$10^3$ GHz (corresponding to 10-100 K in the $k_B$ units). On the other hand, the exchange interaction is weakened at the surface due to reduced number of the nearest neighbors and various irregularities. So the inequality Eq.(3) looks realistic and enables one to separate the total spin system into two quasi-independent subsystems, the bulk and surface ones, and consider the surface-bulk interaction as a perturbation.

Let us consider the bulk in the ferromagnetic (superparamagnetic) state with uniform magnetization and describe it as a single giant spin S>>1. The anisotropy field $B_{av}$ in the bulk is assumed to be relatively small, $B_{av}$<<B, so the magnetic moment of the bulk is directed practically along **B**. By contrast, the surface anisotropy field $B_{as}$>>B; as a result, the surface spins $s_j$ are aligned approximately along the radii, see Fig. 8. Let us calculate now their effect on the EMR frequency of the bulk spins.

Accounting for a great difference between the resonance frequencies of the surface and bulk spins, and employing axial symmetry of the problem, one can keep only z-z terms in the surface-bulk interaction Hamiltonian:

$$H_{sv} = \Sigma_j A_j s_{zj} S_z \qquad (4)$$

Here the summation is over the surface spins $s_j$, and $A_j$ includes both dipole-dipole and exchange interactions.

Our prime interest lies in the superparamagnetic behavior and thus in the rather high temperature range. Let us suppose that the surface layer can be described as a paramagnetic spin system. In such a case, one can use the standard axially symmetric spin-Hamiltonian,

$$H_s = \gamma \sum_j (\vec{B} \cdot \vec{s}_j) + D_s s_{nj}^2 \qquad (5)$$

where $s_{nj}$ is the projection of the j-th surface spin on the local anisotropy axis, **n** is the unit vector perpendicular to the surface, and $D_s \equiv -B_{as}/2s$ is the anisotropy parameter for the surface spins. Suppose at the moment that the eigenvalues of the Hamiltonian, Eq.(5), are not affected substantially by the $J_{sv}$ term. The spin-spin interactions on the surface are not shown explicitly in Eq.(5); we will return to this point later.

Consider, for example, the $Fe^{3+}$ ions (s=5/2). Suppose that $D_s$ is negative, and

$$|D_s| \gg \gamma B \qquad (6)$$

The latter inequality is the central point of our model; obviously, it is consistent with Eq. (3). The energy diagrams for the lowest Kramers doublet (±5/2) at various polar angles θ are shown in Fig. 8. The external magnetic field B strongly splits the doublet at the pole (θ=0), whereas no splitting occurs in the first approximation at the equator (θ=π/2). As a result, the equilibrium polarization of the surface spins, $<s_z(\theta)>$, reaches its maximum at the pole and zero on the equator, being proportional to $\cos^2\theta$.

One can see that the angular distribution of the surface spin polarization is axially symmetric, with the symmetry axis directed along **B**. It should be emphasized that this effect is not associated with any deviation from the spherical shape and should be differentiated from the surface-induced magnetic anisotropy [25]. Instead, the non-uniform magnetization on the surface produces unidirectional field seen by the bulk spins and so leading to temperature-dependent shift of the EMR line to lower fields.

To estimate this effect qualitatively, one can use the mean-field approximation and substitute $<s_z(\theta)>$ for $s_{zj}$ in Eq.(4). At relatively high temperatures, the Curie-Weiss law

$$<s_z(\theta)> \propto (T-\theta_{CW})^{-1} \qquad (7)$$

is expected for $<s_z(\theta)>$, where the Curie-Weiss temperature $\theta_{CW}$ accounts for exchange interactions between the surface spins. Obviously, this leads to similar temperature dependence of the EMR shift. As seen from Fig. 4(a) (inset), this is consistent with the experimental data above $\theta_{CW}$=170K. At lower temperatures, however, this approximation breaks down, and proper accounting is needed for the short-range spin ordering on the surface. An estimation shows that the dipolar magnetic field created by the surface spins in the center of a maghemite spherical particle with R=2.4 nm is about a few Gauss in magnitude This value is considerably lower than the experimental shifts shown in Fig. 4(a), hence the surface-bulk exchange interactions must be taken into account.

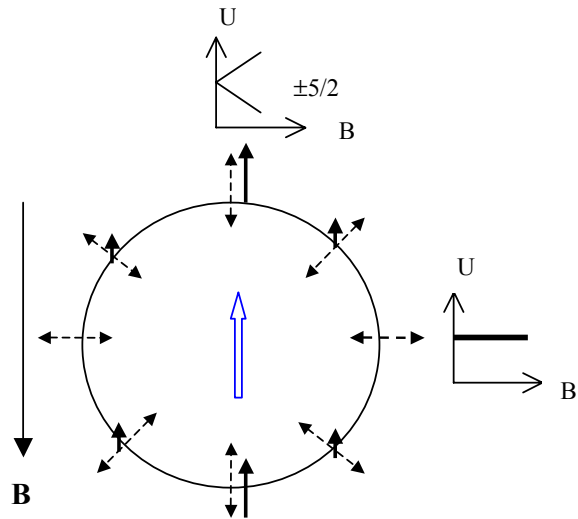

Fig. 8. Surface polarization of a spherical particle. The bulk spin S (the big arrow in the centre), the surface components, $s_z$, (solid arrows), and surface anisotropy axes (dashed double arrows) are indicated. The B-U diagrams show the lowest Kramers doublet at the pole ($\theta=0$) and at the equator ($\theta=\pi/2$).

The analysis shows that in this case the surface-volume interaction cannot be considered as a small perturbation, and required much more sophisticated calculations. One can see, however, that our consideration remains qualitatively valid beyond the restrictions of Eqs.(3) and (6) as well. In any case, strong surface anisotropy results in an increase of the average resonance frequency of the surface spins as compared to $\gamma B$. Due to exchange interaction, this frequency shift (the first spectral moment) is partly transferred to the bulk spin system, leading to the corresponding shift of the EMR spectrum toward lower fields. We shall account for this effect using some phenomenological assumptions, see Section III.E.

Apart from the shift of the resonance line, the broadening of the bulk EMR spectrum would arise due to the surface non-homogeneity and the fluctuating part of the surface magnetization which is proportional to $\langle s_z^2 \rangle - \langle s_z \rangle^2$. One can expect that the temperature dependence of the line broadening would be similar to that of the shift.

### D. The narrow spectral component

Let us discuss now the origin of the narrow feature which is clearly seen at the center of the EMR spectra in both liquid and solid diluted samples at high enough temperatures. As seen in Fig. 5 (a), its amplitude decreases upon cooling approximately obeying the Arrhenius law with the activation energy of about 850 K (in $k_B$ units). On the other hand, the shape and the width of the narrow line are found to be practically unchanged. This suggests that the temperature dependence of the peak magnitude shown in Fig. 5 (a) reflects the behavior of the integrated intensity, at least in the central (peak-to-peak) part of the narrow component.

Berger et al. [13, 14] suggested that the narrow component observed in the EMR spectra of superparamagnetic objects is due to the contribution of very small particles. If $\xi \ll 1$, that would lead to a strong reduction ("dressing") of the anisotropy field and collapse of the spectrum into single Lorentzian line at g=2. Detailed analysis shows, however, that this scenario does not describe our case. According to Refs. [13, 14], the width of the narrow component increases steeply upon cooling due to reduction of the dressing effect [3, 4]; by contrast, the width of the narrow line shown in Fig.

3 remains constant until eventual disappearing at low temperatures. Besides, the Berger-Kliava model [13, 14] cannot explain the nearly exponential temperature dependence of the line amplitude shown in Fig. 5 (a).

Another explanation was proposed by Gazeau et al. [9]. They referred to the RS theory modified by accounting for "inhomogeneous broadening" caused by the dependence of FMR frequency on the angle, $\psi$, between the magnetic moment $\mu$ and magnetic field **B** [ 30, 31 ]. Unfortunately, the detailed theory was not presented in Refs. [9, 30, 31], except for the case of B=0. In the next Section, we suggest a model which, in our opinion, is compatible with this idea. We show that the anisotropy terms are cancelled in the first order at $\psi=\pi/2$, so the narrow component arises at this excited state with the probability of $\exp(-\mu B/k_B T)$.

Whatever the origin of the narrow spectral component may be, it experiences dipolar magnetic fields produced by the particles. The dipole-dipole interaction depends on the distance between the magnetic objects and must lead to broadening of the resonance line with increase in particle concentration. To estimate let us use the statistical theory of the dipole-dipole broadening of EPR in magnetically diluted spin systems [32] which can be readily generalized to our case. According to [32], the Lorentzian line-shape is predicted, with the peak-to-peak width of the first derivative signal:

$$\delta_{pp}^{(1)} = \frac{16\pi^2}{27} Mc \qquad (8)$$

and the full width at the half height of the second derivative central peak:

$$\delta_{1/2}^{(2)} = \frac{16\pi^2}{27\sqrt{3}} Mc \qquad (9)$$

For example, for the highest $c$ of 0.018 presented in Fig. 5 (b), one gets $\delta_{1/2}^{(2)} \cong 24$ G, whereas the experimentally observed concentration dependent part of the broadening does not exceed 3-4 G. This discrepancy shows that the dipolar fields contributes mostly to unobservable distant wings rather than to the peak-to-peak region, and the real shape of the dipolar broadened line is not Lorentzian. Such case is characteristic of non-random space distribution of magnetic entities, with a tendency to aggregation. In the aggregates, the dipolar fields are much stronger than at mean distances at a given $c$. This results in transferring the resonance absorption to distant line wings, accompanied by corresponding drop in the central region, in accordance with Fig. 5 (b). In other words, the central (narrow) part of the line is provided only by the "free" particles which are not coupled in dipolar clusters. The role played by particle aggregation will be discussed in more detail in Section III.F.

**E. Fitting the EMR spectrum**

In this Section, we make an attempt to fit, at least qualitatively and on phenomenological level, the shape of the observed EMR spectra, including characteristic temperature evolution of both the narrow and broad components. The simultaneous existence of two distinct spectral features is difficult to explain in the frames of the RS theory [3, 4] which suggests averaging over all possible states due to fast rotary diffusion of the magnetic moment. To overcome this problem, we assume that thermally activated jumps of $\mu$ between different orientations are not fast enough, and its Larmor precession can be definitely distinguished at each particular angle $\psi$ between $\mu$ and **B**. In such a case, the observed FMR spectrum can be represented as a sum of the signals corresponding to various $\psi$, with proper account for their probabilities.

It should be emphasized that this approach is applied only to bulk spins undergoing uniform precession. It is known that the maghemite magnetic structure is typical of ferrites and contains sublattices with both positive and negative exchange interactions. Nevertheless, the standard FMR picture of the uniform precession is fully applicable to the net magnetic moment (with exception of a few special points and very high external fields) [33]. The effect of the surface is taken into account phenomenologically, as a perturbation to the idealized consideration of the core.

The bulk anisotropy of $\gamma$-$Fe_2O_3$ is cubic, with $K_c \sim -4.7 \cdot 10^3$ J/m$^3$ [34] which corresponds to the anisotropy field of about 30 G. This value is much less than the observed width of the EMR spectrum and can be neglected. Instead, we introduce an axial anisotropy field $B_a$ as a fitting parameter. It can be caused by small deviations from the spherical shape as well as the remains of the surface anisotropy averaged over the volume through strong exchange interactions. In what follows, the high-field approximation,

$$B \gg B_a \qquad (10)$$

is adopted, so the direction of the effective field (the precession axis) is practically coincides with **B**.

We start with the well-known expression for the energy of an anisotropic ferromagnet in a magnetic field:

$$U = -(\boldsymbol{\mu} \cdot \mathbf{B}) - KV\cos^2\varphi \qquad (11)$$

where $K = B_a M/2$ is the specific anisotropy energy and $\varphi$ is the angle between $\boldsymbol{\mu}$ and the anisotropy axis **n**. Applying the standard master equation for the classical magnetic moment and neglecting the relaxation, it can be shown [35] that the precession of $\boldsymbol{\mu}$ under condition of Eq.(8) occurs at the frequency

$$\omega_{\psi,\theta} = \omega_0 + \gamma B_a \cos\psi P_2(\cos\theta) \qquad (12)$$

where $\omega_0 = \gamma B$; $P_2(y) = (3y^2-1)/2$ is the Legendre polynomial, and $\theta$ is the angle which **B** makes with the anisotropy axis. The expression of Eq.(12) differs from the standard FMR one by the factor of $\cos\psi$ which accounts for deviation of $\boldsymbol{\mu}$ from its ground-state direction ($\psi=0$) due to thermal excitation and describes the superparamagnetic behavior.

We consider this frequency as the center of the individual resonance line ("spin packet") which is related to the given value of $\psi$. The corresponding resonance magnetic field reads:

$$B_{\psi,\theta} = B_0 - B_a \cos\psi P_2(\cos\theta) \qquad (13)$$

where $B_0 = \omega/\gamma$, and $\omega$ is the microwave frequency.

At this stage, let us switch to the notations commonly used in the field of EPR. Namely, we denote $\cos\psi = -m/S$, where m is the magnetic quantum number determining the z-projection of the total spin on the **B** direction. Further, let $\gamma B_a = -2DS$. In these notations, Eqs.(11)-(13) become very similar to the equations related to the paramagnetic spin system described with the standard axially symmetric spin Hamiltonian (compare to Eq.(5)):

$$\hat{H} = \gamma(\vec{B}\cdot\vec{S}) + DS_n^2 \qquad (14)$$

In the high-field approximation, the eigenvalues of Eq.(14) are [32]:

$$E_{m,\theta}/\hbar = \gamma mB + Dm^2 P_2(\cos\theta), \qquad (15)$$

and the corresponding resonance field reads:

$$B_{m,\theta} = B_0 + (2m+1)DP_2(\cos\theta) \qquad (16)$$

Comparing Eqs. (15), (16) with Eqs. (11), (13), one can see that the "quantization" description fully coincides with "ferromagnetic" formalism at $S, m \gg 1$. Obviously, in our case ($S \sim 10^3$) the energy levels with different $m$ should by considered as a continuous band.

This similarity clearly points to the intermediate position occupied by superparamagnetic nanoparticles between the ordinary ferromagnets and exchange coupled paramagnetic clusters (or molecular magnets). In the ferromagnetic case, the spin system always remains in its ground state ($\psi = 0$); in the exchange cluster, the situation corresponds to the high-spin ground multiplet separated from the upper one by a gap of about $JS \gg k_BT$. Since the spin wave energy is inversely proportional to the particle diameter [36], this picture can be well justified for our nanoparticles. Note that similar approach was employed previously when interpreting magnetization [37, 38] and nuclear spin relaxation [39] in superparamagnetic samples. A model with discrete electronic states was also introduced to describe the ferromagnetic metallic nanograins [40].

In our calculation, we should take into account the allowed ($m$, $m+1$) transitions with the probabilities [32]

$$W_{m,\theta} = Ag(B - B_{m,\theta})[S(S+1) - m(m+1)] \quad (17)$$

where A is a proportionality factor and $g(B-B_{m,\theta})$ is the form-factor of the resonance line at the transition involved. In what follows, either Lorentzian or Gaussian shapes were tested, with the corresponding value of the half-width $\delta$ as a fitting parameter. Finally, one should take into account the equilibrium (Boltzmann) distribution of populations on the magnetic sublevels,

$$\rho_{m,\theta} = Z_\rho^{-1} \exp\left(\frac{-E_{m,\theta}}{k_BT}\right) \quad (18)$$

where $Z_\rho$ is the partition function. The resonance absorption is proportional to the population difference at the adjacent levels, $\rho'_{m,\theta} = \partial \rho_{m,\theta} / \partial m$. As a result, the shape of EMR absorption spectrum for an assembly of nanoparticles with random distribution of the crystalline axes reads:

$$G(B - B_0) = \int_0^\pi \sin\theta d\theta \int_{-S}^{S} W_{m,\theta} \rho'_{m,\theta} dm \quad (19)$$

Before passing to the comparison with the experimental data, let us discuss distinctions between the approach described above and the RS theory [3, 4]. In the latter case, averaging is performed over all $\psi$ values, suggesting the validity of the Landau-Lifshits (LL) equation for the magnetic nanoparticle as a whole. This means, in particular, that the longitudinal and transversal relaxation times ($T_1$ and $T_2$) are assumed to be equal and determined by the LL relaxation parameter. Unlike this, Eqs. (17)-(19) include the summation over all ($m$, $m+1$) transitions, assuming that each of them is well-distinguished and characterized by specific $T_1$ and $T_2$ values, according to the Bloch equations. This suggestion is supported by the measured value of $T_1 = 10$ ns, which exceeds considerably the value of $T_2 \sim 2$ ns estimated from the width of the narrow feature in diluted samples. Note that the value of $T_2$ becomes shorter in the concentrated ferrofluid and inside the aggregates which can be formed due to interparticle dipole-dipole interactions.

The "quantization" approach described above results in pronounced change in calculated spectra as compared to the RS model. In particular, the appearance of the narrow spectral component is predicted. The narrow component results from the contribution of the states in the vicinity of $\psi=\pi/2$ which are not affected by the anisotropy field (see Eqs. (13), (16)), and thus not broadened by random distribution of the symmetry axes. Since the energy of these states lie well above of the ground level, the intensity of the narrow feature decreases exponentially upon cooling. Note that a similar thermally activated double-pattern spectrum has been observed, as well, in very different systems of magnetically dilute $LaGa_{1-x}Mn_xO_3$ at intermediate Mn concentrations, and was ascribed to Mn clusters [41]. This similarity in EMR spectra and temperature behavior may have significant physical meaning, revealing continuous transfer from the exchange coupled spin clusters in diamagnetic host lattice to superparamagnetic nanoparticles.

Our simulations show that good agreement with the experiment cannot be achieved if the line-width $\delta$ is kept independent on $m$. To get better fitting of the experimental spectrum and, in particular, to describe its broadening and shifting to lower fields with decrease in temperature, we assume that the individual width and resonance field of a given transition depend on its quantum number $m$ as

$$\delta(m) = \delta_0 + a|m| \qquad (20)$$

$$B'_{m,\theta} = B_{m,\theta} + c|m| \qquad (21)$$

where $\delta_0$, $a$ and $c$ are phenomenological fitting parameters. This suggestion may be qualitatively justified by the surface-volume interaction Hamiltonian (Section III,C) which includes the $\hat{s}_z \hat{S}_z$ operators and contains matrix elements proportional to $m$. More detailed theory is now in progress and will be published elsewhere. Using Eqs.(20) and (21), we fit the experimental data shown in Fig. 9 .

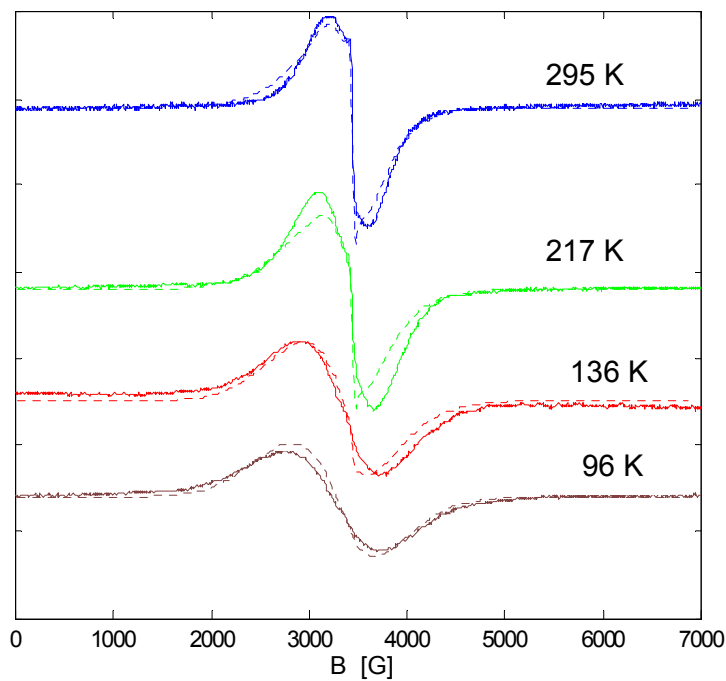

Fig. 9. Fitting of the experimental data (nanoparticles in polymer matrix) with Eqs.(16)-(21). Solid traces: experiment, dotted traces: theory. Temperatures are indicated in the figure.

The parameters employed at this fitting were: η=800K; $B_a$ = 700 G; $a$ = 2.8 G; $c$ = 0.41 G; the Gaussian line shapes was used as $g(B_{m,\theta}-B_0)$. According to the observed concentration dependence of the narrow component (Figs. 2 and 5(b)), the value of $\delta_0$ = 30 G was ascribed to "free" particles, whereas $\delta_0'$=200 G was used for the aggregates. The fraction of the free particles was assumed to be of 20%. As one can see, the calculated spectra agree qualitatively with the experimental ones.

**F. Effect of field freezing**

Let us discuss now the field freezing (FF) experiments (Section II, Figs. 6, 7). Our data are consistent with those obtained previously by Gazeau et al. with maghemite nanoparticles in anionic ferrofluid [8]. In both cases, the specific angular dependence of the resonance field after the FF procedure was observed. As it is seen from Fig. 6, this dependence has only one maximum in the range of 0 - 180 deg and can be well fitted by simple expression $\cos^2\beta$. As it was mentioned by Gazeau et al. [8], such behavior evidences for uniaxial anisotropy, in contrast with the cubic symmetry known for the bulk γ-$Fe_2O_3$ material. It was suggested [8] that this axial anisotropy is related to individual nanoparticles and caused by the surface effects combined with deviation from the spherical shape, as it was predicted by Néel [25]. Consider this assumption in more detail.

According to Ref. [4], the orientational distribution f ($n$, $B_{fr}$, $T_{fr}$) of the anisotropy axes under FF conditions results from a competition between the magnetic and anisotropic energy of a nanoparticle with the thermal energy ($k_BT$) as:

$$f(\vec{n}, \vec{B}_{fr}, T_{fr}) = Z^{-1} \int exp\left[\frac{\mu}{k_B T_{fr}}(\vec{e}\cdot\vec{B}_{fr}) + \frac{K}{k_B T_{fr}}(\vec{e}\cdot\vec{n})^2\right] d\vec{e} \qquad (22)$$

Here **n** and **e** are the unit vectors in the directions of the easy axis (making an angle θ with $B_{fr}$) and the magnetic moment **μ**, respectively; $K \equiv \mu B_a/2$ is the anisotropy constant, and Z is the normalizing factor. Explicit formulas for $f$ (θ, $B_{fr}$, $T_{fr}$) and Z are given in Ref. [4].

Following the idea of Ref. [42], Gazeau et al. [8] employed a simplified method to calculate the shift of the resonance line, and obtained the following expression for the difference between the line positions at $B_m \perp B_{fr}$ (β = 90 deg.) and $B_m \parallel B_{fr}$ (β = 0):

$$B_0(90) - B_0(0) = \frac{3}{2}\langle\Delta\rangle \equiv \left(\frac{3}{2}\right)\cdot 4\pi \int_0^{\pi/2} f(\theta, B_{fr}, T_{fr})\Delta(\theta)Sin\theta d\theta \qquad (23)$$

where Δ(θ) is the shift of the resonance position for an individual nanoparticle determined by the magnitude and orientation of the anisotropy field as

$$\Delta(\theta) = B_0 - B_{res} = B_a P_2(\cos\theta) \tag{24}$$

This method, though not rigorous, can be employed when the shape of the EMR spectrum is nearly symmetric. To determine two unknowns, μ and $B_a$, we used the experimental dependence of $B_0(90^0)$-$B_0(0)$ on $B_{fr}$, see Fig. 7. The data were fitted using Eq.(23) with $T_{fr}$ = 180 K; the best fit is shown by the solid curve. Parameters determined from this fitting procedure are: η = (4300±400) K and $B_a$ = (800±100) G. Though the obtained value of $B_a$ is close to the upper limit determined by the observed width of the EMR spectra, the η value disagrees dramatically with all other experimental data, such as the particle size and temperature dependence of susceptibility.

One can try to use a realistic value of η= 800 K and then resolve Eq.(23) at high enough $B_{fr}$, such as 7 kG or 70 kG. The calculated curves (see the dotted and dashed lines at Fig. 7) disagree strongly with experimental data; besides, this fitting yields $B_a$ > 2000 G, far beyond of the acceptable upper limit. Direct calculations of the EMR spectrum after the FF procedure with either the RS model [3, 4] or the method described in Section III.E did not provide any satisfactory agreement with the experimental data at any combinations of the μ and $B_a$. Note that the same problem arises if one analyses the data of Gazeau et al. [8, 9]. In fact, the reported value of $B_a$= 2.2 kG for d= 4.8 nm [8] should lead to extremely asymmetric and very broad spectrum, in strong disagreement with the experimental data.

Thus, the model of alignment based on Eqs. (22)-(24) does not provide a good quantitative description of experimental data on the surface functionalized maghemite nanoparticles: in fact, the alignment achieved in moderate magnetic fields, $B_{fr}$~4-7 kG, is significantly enhanced as compared with theoretical predictions based on the single-particle anisotropy. Note that the best-fit value of η corresponds to the 5-6 times enlarged magnetic moment of the particle, clearly pointing to some collective effects (aggregation) caused by interparticle interactions.

The phenomenon most likely affecting the results of FF experiments is the formation of chains aligned in the **B** direction. Such chains were observed previously in various magnetic fluids (see, for example, Ref. [43]). Note that our nanoparticles are separated by an organic nanolayer which prevents close contact and the exchange interaction between them. Only magnetic dipole-dipole interaction should be taken into account. The dipolar magnetic field seen by a particle situated in the middle of a chain parallel to **B** reads:

$$B_{dip} = \frac{\mu_0}{4\pi} \sum_i \frac{MV(3\cos^2\theta - 1)}{r_i^3} \tag{25}$$

where $r_i$ is the distance between the centers of the given particle and its *i*-th neighbor and θ is the angle which **B** makes with the chain axis. Here the high-field approximation is accepted where all magnetic moments are directed along **B**. Note that the interparticle dipolar interaction can be considered as small perturbation to the main Hamiltonian and so the foregoing consideration remains valid.

To estimate the line shift under the FF procedure, we limit the sum, Eq.(25), by the nearest neighbors (the subscript *i* will be omitted) and suppose

$$r = d + \varepsilon \tag{26}$$

where d is the particle diameter. Substituting d = 5 nm, M =4·$10^5$ A/m, and assuming $B_{dip}(\theta=0)$ = $B_0$-$B_0(0)$ = 310 G (see Figs. 6, 7), one gets a realistic value of ε = 2 nm. Assuming the direction of the chain to be frozen along $B_{fr}$, the $\cos^2\beta$ dependence (Fig. 6, inset) can be readily obtained.

Note that the process of chain formation and their alignment in magnetic field is not strictly described with Eq.(22) intended for a single anisotropic particle. Nevertheless, the obtained value of η~4300 K enables us to estimate, in a crude approximation, the mean chain length of 5-6 particles.

Note that, according to Eqs.(25) and (26), the value of $B_{dip}$ is proportional to $[d/(d+\varepsilon)]^3$ and decreases with decreasing the particle diameter. This agrees with the experimental data reported by Gazeau et al. [8].

## IV. Conclusion

In conclusion, EMR spectra and longitudinal spin relaxation have been studied in both concentrated magnetic fluid and diamagnetically diluted objects containing 5 nm maghemite nanoparticles. The spin-relaxation time $T_1$ is found to be in the order of 10 ns and temperature independent at T=77-300 K. EMR spectra consist of broad and narrow components, the latter having temperature independent g-factor and revealing thermally activated intensity with $E_a/k_B \cong 850$ K. Unlike this, the broader component shows considerable broadening and shifting to lower fields upon cooling. This behavior, as well as almost symmetrical shape of the broad line, cannot be satisfactory described by the standard RS theory [3, 4]. It is shown that strong radial anisotropy experienced by surface spins in combination with the effect of external magnetic field **B** leads to the axially symmetric distribution of spin polarization even for spherical particles, which may be partially responsible for the shift and broadening of the observable EMR spectrum through the surface-bulk spin-spin interaction.

To explain an appearance and nearly exponential temperature dependence of the narrow spectral component at g=2, the hypothesis is proposed of independent contributions of the energy states differing by the ψ. At right angle between **μ** and **B**, the anisotropy-related inhomogeneous broadening disappears, thus producing a narrow spectral component with a probability of ~ $\exp(-\mu B/k_B T)$. This "quantization" approach enables one to fit, at least qualitatively, the shape of the observed spectrum in the whole temperature range studied (77-295 K).

Experiments on freezing of the liquid samples in magnetic field $B_{fr}$ (the FF procedure) provide evidence for axial anisotropy in the resulting EMR spectrum. Quantitative analysis showed that these data cannot be described by the single-particle axial anisotropy. Instead, the interparticle dipole-dipole interactions leading to particle clustering should be taken into account. The model of a chain parallel to **B** and consisting of 5-6 particles is shown to be consistent with the experimental FF shift and angular dependence. The minimum interparticle distance caused by the organic covering layer is estimated as to be around 2 nm. The tendency to particle aggregation is also confirmed by the dependence of the narrow line parameters on diamagnetic dilution.

**Acknowledgments**

The work was partly supported by National Science Foundation (NSF) CREST Project HRD-9805059, Russian Foundation for Basic Research (Grant 05-02-16371), and Program P-03 for Basic Research of Russian Academy of Sciences.

The authors are greatly indebted to F.S. Dzheparov for the help in theoretical calculations and fruitful discussions and R. R. Rakhimov and V. V. Demidov for the help in EMR experiments.